\documentclass{mem}
\usepackage{natbib}\usepackage{txfonts}\usepackage{balance}
\usepackage{graphicx}
\usepackage{epstopdf}
\usepackage[a4paper,breaklinks,dvipdfm]{hyperref}
\idline{75}{282}
\begin{document}
\def\teff{$T\rm_{eff }$}
\def\kms{$\mathrm {km s}^{-1}$}
\newcommand{\xss}{XSS1227\,}
\newcommand{\fgl}{2FGL1227\,}

\title{
The peculiar source XSS J12270-4859: a LMXB detected by FERMI ?
}

   \subtitle{}

\author{
J.M. \,Bonnet-Bidaud\inst{1} 
\and D. \, de Martino\inst{2}
\and M. \, Mouchet\inst{3,4}
\and M. \, Falanga\inst{5}
\and T.~Belloni\inst{6}
\and N.~Masetti\inst{7}
\and K.~Mukai\inst{8}
\and G.~Matt\inst{9}
          }

  \offprints{J.M. Bonnet-Bidaud, \email{bonnetbidaud@cea.fr}}

\institute{
CEA	Saclay, DSM/Irfu/Service d'Astrophysique, F-91191 Gif-sur-Yvette, France
\and INAF - Osservatorio Astronomico di Capodimonte, I-80131 Napoli, Italy
\and Laboratoire APC, Universit\'e Paris-Diderot, F-75013 Paris, France
\and LUTH, Observatoire de Paris, CNRS, Universit\'e Paris-Diderot, F-92190 Meudon, France
\and International Space Science Institute (ISSI), CH-3012 Bern, Switzerland
\and INAF-Osservatorio Astronomico di Brera,  I-23807 Merate, Italy
\and INAF Istituto Astrofisica Spaziale, I-40129, Bologna, Italy
\and CRESST and X-Ray Astrophys. Lab., NASA Goddard SFC, Greenbelt, MD 20771, USA 
\and Dipartimento di Fisica, Universit\'a Roma III, I-00146, Roma, Italy
\\
}

\authorrunning{J.M. Bonnet-Bidaud }

\titlerunning{The Peculiar Source XSS J12270-4859}

\abstract{
The X-ray source XSS J12270-4859 has been first suggested to be a magnetic cataclysmic variable 
of Intermediate Polar type on the basis of its optical spectrum and a possible 860 s X-ray periodicity. 
However further X-ray observations by the Suzaku and  XMM-Newton satellites did not confirm this periodicity
but show a very peculiar variability, including moderate repetitive flares and numerous absorption dips. 
These characteristics together with a suspected 4.3\,h orbital period would suggest a possible link with 
the so-called "dipping sources", a sub-class of Low-Mass X-ray Binaries (LMXB). Based on the released 
FERMI catalogues, the source was also found coincident with a very high energy (0.1-300 GeV) VHE source 2FGL J1227.7-4853.  
The good positional coincidence, together with the lack of any other bright X-ray sources in the field, makes 
this identification highly probable. However, none of the other standard LMXB have been so far 
detected by FERMI. Most galactic HE sources are associated with rotation-powered pulsars.
We present here new results obtained from a 30\,ksec high-time resolution XMM observations 
in January 2011 that confirm the flaring-dipping behaviour and provide upper limits on fast X-ray pulsations. 
We discuss the possible association of the source with either a microquasar or an accreting rotation powered pulsar.
}
\maketitle

\section{The identification of XSS J12270-4859}
The X-ray source XSS J12270-4859 (hereafter \xss) was discovered from the Rossi XTE survey 
by \citet{saz04}.   
 It was subsequently detected at energy higher than 20 keV from observations by the INTEGRAL
  satellite \citep{bird10}. An identification campaign leds to its identification with an (m$_v \sim16$) optical star 
  which shows Balmer emission lines typical of an accreting white dwarf \citep{mas06}. 
  Pointed observations by RXTE seemed further to consolidate the suggestion of a cataclysmic variables 
  of Intermediate Polar (IP) when 860 sec X-ray modulation was reported, typical of IP rotations \citep{but08}.
However, fast optical photometry failed to confirm these pulsations \citep{pre09}.
Furthermore, the source was more extensively observed in X-rays by the Suzaku and XMM satellites 
\citep{sai09,ddm10},  both observations ruling out the presence of significant X-ray pulsations. 
A complete account of the 2009 XMM observations is given in \citet{ddm10} (hereafter DDM10).
The detailed observations show the source with a power-law X-ray energy distribution already
more suggestive of a low-accreting X-ray binary (DDM10, \citet{fal10}) but the discovery 
of a probable association with a high energy (E $>$ 100 MeV) gamma-ray source makes it 
a very peculiar object.

\section{The association with the FERMI source}
Analysing our January 2009 XMM observations, we were the first to note 
the possible association of the source with a high energy counterpart.
The first FERMI catalogue was released in Jan. 2010. The point source catalogue of the
high-energy (100 MeV to 300 GeV) gamma ray sources detected by the 
Large Area Telescope (LAT) is based on observations of the first 11 months of 
the science mission. It lists a significant source 1FGL J1227.9-4852
 at a distance less than 2 arcmin from \xss with an associated uncertainty 
of 3.6 and 6.0 arcmin respectively for a 68 \% and 95\% confidence. 

A refined position was recently provided by the second-year FERMI catalogue \citep{abdo11} 
with the source, now labeled 2FGL J1227.7-4853 (herafter \fgl).
\fgl is detected at a significance of 24.3\,$\sigma$ 
with a  (0.1--100\,GeV) flux of 3.38$\pm 0.22 \times 10^{-11}\,\rm 
erg\,cm^{-2}\,s^{-1}$ and a best fit power law photon index of 2.33$\pm$0.08 with a high energy cut-off around 4 GeV.  
It is at a separation with respect to \xss of 2.7 arcmin with a 95\% confidence uncertainty of 3.6 arcmin.
The source \xss is the only bright X-ray source in a (6x6) arcmin field centered around \fgl 
making the association quite reliable.
The \fgl position is consistent with the one derived by \citet{hill11} who also conducted 
a radio survey of the error box. Three faint sources were found in the box.  One is coincident 
with \xss (to within 1 arcsec) with a flux density of 0.18 mJy at 5.5 GHz.

\begin{figure*}[t!]
\resizebox{\hsize}{!}{\includegraphics[clip=true, angle=0]{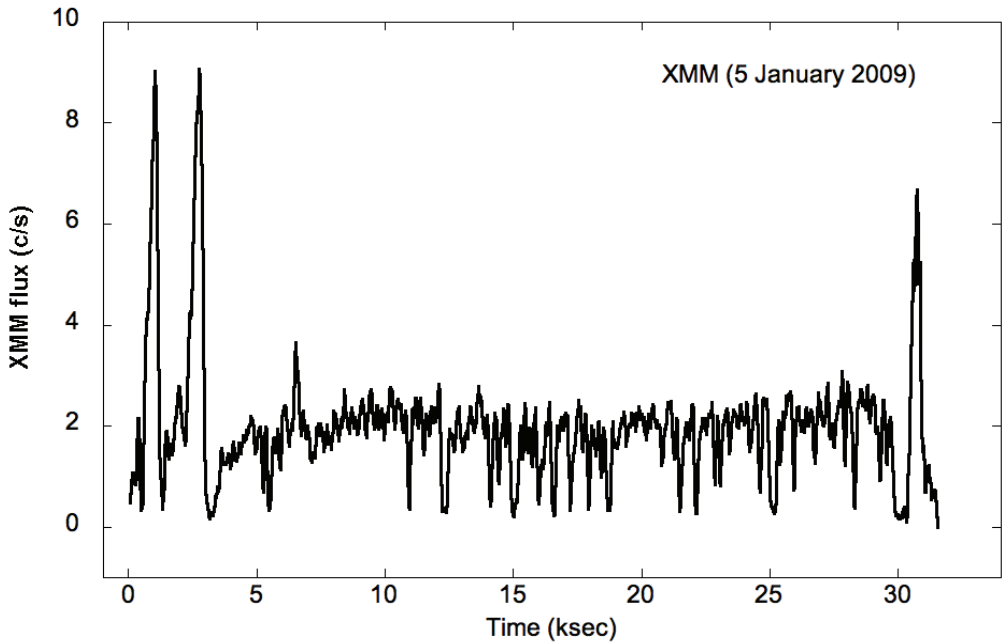}}
\caption{\footnotesize
The (0.2-10) keV light curve of \xss obtained with XMM on 5 January 2009. Repetitive absorption  dips are clearly visible all through the observations with also three large ($\sim$ 5-10 min.) flares.
}
\label{lcurve}
\end{figure*}

\section{The X-ray variability of XSS J12270-4859}
The atypical X-ray variability of \xss was first uncovered by Suzaku observations \citep{sai09} 
that revealed the presence of repetitive flares and numerous absorption dips with significant spectral changes.
However, numerous gaps in the Suzaku data prevented an accurate study.
A first continuous 35\,ksec XMM observation was obtained on January 5, 2009 supplemented by 
additional optical photometric data (DDM10).
Figure \,\ref{lcurve} shows  the (0.2-10 keV) XMM light curve
with the presence of three large flares of typical duration of $\sim$ 10 min. and 
more than 20 absorption dips with very small residual flux. 
This same type of behaviour was consistently seen for the source when additional RXTE \citep{sai11}
and XMM observations \citep{ddm12} were obtained.

New XMM observations obtained on January 1st, 2011 reveals again
more than 35 absorption dips; most of the dips being short ($\la$ 3 min.) 
but at least three long dips are also visible with duration more than 10 min. 
The dips are reminiscent of the dipping LMXB sources \citep{dia06}, 
such as EXO\,0748-676 \citep{bobi01}, 
but in these sources they occur at specific orbital phases and with a spectral hardening that is not 
observed in \xss (see below).
 Three ($\sim$ 5-fold) flares with duration $\sim$ 5-10 minutes are also present at the end of the observations. 
 These flares are similar to those observed previously.
 Showing much longer timescales and a rather a-typical "inverted" shape 
 with a slow rising and a sharp drop, they are quite different from the typical bursts seen 
 among other X-ray binaries.

The broad-band (0.2-100\,keV) XMM and INTEGRAL combined spectrum of \xss\, is found featureless 
and well described  by a  weakly absorbed power  law with a photon index of $\Gamma_{ph}$ $\sim 1.7$ 
and an hydrogen column density  $\sim 1\times  10^{21}\rm cm^{-2}$ consistent with the interstellar extinction. 
The source does not show any significant Fe line commonly seen in accreting white dwarfs.
There is no apparent cut-off up to 100 keV and the derived X-ray luminosity is 
$\rm L_X \sim 5\times 10^{33}\,(d/1kpc)^{2}\,\rm erg\,s^{-1}$.
Time-resolved X-ray spectroscopy also  clearly shows that the flares and dips in general are accompanied 
by no significant spectral change. However the flares are usually followed by a dip than can be 
described with severe absorption
$\rm N_{H} \geq 6.1\times 10^{22}\,cm^{-2}$ (see DDM10).

There is at present no definitive measure of the orbital period of the system. 
The best indication has come so far from our V and J optical photometry that shows a significant modulation 
at a period of $\sim$ 4.3 h. A marginal evidence of a low (4\%) amplitude variability at this period is also 
found in the X-ray and UV ranges (DDM10). This periodicity has still to be confirmed  by upcoming
optical spectroscopy but seems to be a good indication of a LMXB.

\section{The high energy spectrum}
\begin{figure}[]
\resizebox{\hsize}{!}{\includegraphics[clip=true,angle=-90]{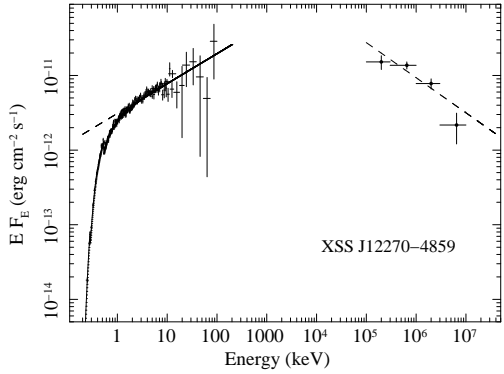}}
\caption{
\footnotesize
The composite XMM-Integral-Fermi energy density spectrum (from \citep{ddm10}).
}
\label{spec}
\end{figure}

The association with \fgl means that the source has an emission extending up to the GeV range 
rather atypical for LMXBs.  
The combined XMM-INTEGRAL-FERMI spectrum is shown in Fig. \,\ref{spec} as 
Energy Density Spectrum (SED) to better show where the peak energy is released.
With a power-law index changing from 1.7 (X-rays) to 2.3 (VHE), the maximum flux will be expected around 1 MeV.
There is a significant energy released above 100 MeV with a ratio 
$\rm L_X$(0.2-100keV)/$L_{\gamma}$(0.1-100GeV)=$0.9$.
The FERMI light curve shows that the source is persistent at VHE with no significant flaring activity 
to within the statistical uncertainty.
Very few X-ray binaries have been so far detected by FERMI. In the last two-year catalogue, 1873 sources 
are detected of which 127 can be considered as being firmly identified and 1170 as being reliably 
associated with counterparts. Among the firm identifications, 83 are pulsars, 28 are AGN, 6 are SNR, 4 are HMXB, 
3 are Pulsar-wind nebulae, 2 are normal galaxies, and one is a nova \citep{abdo11}. 
Among the HMXBs, all sources are long-period (P$> 3$d) massive systems apart from 
the microquasar candidate Cyg X-3 (P=0.2d) that is only transitorily detected.
Formal associations with low- mass X-ray binaries are reported for three 2FGL sources 
but all three are located in globular clusters, and the observed emission can be readily 
explained by the combined emission of millisecond pulsars. 
\fgl could be therefore the first LMXB emitting at high energies.

\begin{table}
\caption{Source comparison}
\label{pulsar}
\begin{center}
\begin{tabular}{lrr}
\hline
\\
Source  & PSRJ1023 & XSS1227 \\
\hline
\\
Pulse           & $1.67$ ms       & ? \\
Orbit             & $4.75$ hr        & $4.3 $ hr ? \\
\hline
Radio Flux  & $ 6$ mJy          & $0.2$ mJy\\
Optical\,(V)   & $15-17.4$       & $15.8-17.3 $ \\
\hline
X-ray flux  &$ 4.6\,10^{-13}$ & $4.5\,10^{-11}$ \\
Spectrum  (PL)& $ 1.3-1.7 $   &  $ 1.7$ \\
Pulsation  &Yes ($11\%$)     &  ($<30\%$) \\
\hline
$\gamma$ flux  &$ 5.5\,10^{-12}$ & $4.1\,10^{-11}$ \\
Spectrum  (PL) & $ 2.9 $   &  $ 2.3$ \\
L$\gamma$\,(1kpc) &  $ 6.5\,10^{32}$ &  $ 4.9\,10^{33}$  \\
\hline
Ratio L$\gamma$/$L_X$ & $ 12 $   &  $ 0.9 $ \\
\hline
\end{tabular}
\end{center}
\end{table}

\section{Discussion}
The nature of \fgl is puzzling. 
The identification of \xss with \fgl appears quite reliable since the absence of other significant 
X-ray as well as strong radio sources in the field evidently excludes an alternate association with another accreting 
binary or a background extragalactic source.
The X-ray source is consistent with an  X-ray binary accreting at a rather low rate
 if placed at a reasonable distance less than a few kpc.
At such distance, the optical magnitude excludes a massive companion and the measured (4.3h) optical modulation
points instead towards a typical LMXB. The key question is therefore the origin of the VHE emission. 
A suggestion has been made by \citet{sai11}, that \xss might be a microquasar accreting at very low rate
with a synchrotron jet. The weak point is however that among the known microquasars,
 such as GRS 1915+105 or XTE J118+480, they usually show an X-ray variability much higher 
 than \xss and none emits at very high energy. 
The only related system, Cyg X-3, is only  very transitorily detected by FERMI in specific conditions 
after a major radio flare and with a significant VHE flux of 
$\rm L_\gamma \sim 3\times 10^{36}\,(d/7kpc)^{2}\,\rm erg\,s^{-1}$.

In view of the content of the Fermi catalogue and the high level of detection of rotation-powered pulsars, 
a more promising system will be a millisecond-pulsar in a LMXB such as PSR J1023+0038
as already proposed by \citet{hill11}.
Interestingly enough, this source was also initially classified as a CV candidate on the basis 
of an emission line optical spectrum spatially coincident with a radio source \citep{bond02}. 
Its X-ray and optical variability and line velocities clearly show an orbital period of $P_{orb}= 4.75$\,hr
but a significant change in accretion was also observed around 2002 when the optical spectrum 
turns to an absorption spectrum \citep{tho05}.

The discovery of a fast (1.6 msec) millisecond pulsar in the system together with the possible 
detection of the pulsation in X-rays suggested that, after an accretion episode, the system might 
have turned to quiescence with a spectrum dominated by the pulsar emission \citep{arch10}.
Unexpectedly, the source was detected by FERMI with
a  (E$>100$MeV) flux of 5.5$\pm 0.9 \times 10^{-12}\,\rm 
erg\,cm^{-2}\,s^{-1}$ and a power law index of 2.9$\pm$0.2 
with gamma-rays  originating either from the pulsar magnetosphere 
or a shock where accreting material interacts with the pulsar wind \citep{tam10}. 
The similarity with \xss is surprising and summarized in Table \,\ref{pulsar}.
The two sources have similar orbits and optical flux and also very similar X-ray 
and gamma ray spectral shapes with only differences in flux, \xss being significantly 
brighter in X-rays with a lower L$\gamma$/$L_X$ ratio. 

PSR J1023+0038 is the first and only known rotation- powered MSP in a quiescent LMXB.
The position of \xss, at 14$^{\circ}$ high above the Galactic plane also favoured the MSP
hypothesis which could be proved by the detection of millisecond pulsations. 
To this purpose, we asked for high-time resolution mode in our 2011 XMM observations 
to search for coherent signals in the frequency range 0.5-1000\,Hz. 
Once corrected for the expected orbital smearing,  no significant pulsations were found 
with an upper limit varying from 15 to 30\% that is still well above the typical 11\% fraction 
seen in PSR J1023+0038. Negative pulsation radio searches were also reported by \citet{hill11} 
that could also be heavily hampered by orbit smearing so that no definitive conclusion 
can yet be drawn. 

The observed FERMI light curve, as well as our last 2011 XMM observations, confirms that \xss is
a rather stable X-ray and VHE source, probably accreting at a very low level. 
It is tempting to relate the VHE emission to the 
particles accelerated by a MSP pulsar as in the pulsar wind scenario, where $\gamma$-rays are generated 
by synchrotron radiation of the electrons (or possibly protons) accelerated in the shock with the 
infalling matter \citep{tak09}. 
This scenario more naturally applied to PSR J1023+0038 where accretion may have ceased
but could also possibly operate when residual accretion exists such as in \xss.

\begin{acknowledgements}
DdM acknowledges financial support by ASI INAF Contract N. I/009/10/0 
\end{acknowledgements}

\bibliographystyle{aa}

\begin{thebibliography}{}

\bibitem[{Abdo et al. (2011)}]{abdo11} 
Abdo A. et al. 2011, arxiv.org/abs/1108.1435

\bibitem[{Archibald et al. (2010)}]{arch10} 
Archibald A. et al. 2010, \apj, 722, 88

\bibitem[{Bird et al. (2010)}]{bird10} 
Bird A. et al. 2010,  \apjs, 186, 1

\bibitem[{Bond et al. (2002)}]{bond02} 
Bond H. et al. 2002,  \pasp, 114, 1359

\bibitem[{{Bonnet-Bidaud} {et~al.}(2001)}]{bobi01}
{Bonnet-Bidaud} J.-M. et al. 2001, \aap, 365, 282
  
  \bibitem[{{Butters} {et~al.}(2008)}]{but08}
{Butters} O.~W. et al. 2008, \aap, 487, 271
  
\bibitem[{Diaz Trigo et al. (2006)}]{dia06} 
Diaz Trigo M. et al., 2006, A\&A 445, 179Ð195 

\bibitem[{de Martino et al. (2010)}]{ddm10} 
de Martino D. et al. 2010,  \aap, 515, 25

\bibitem[{de Martino et al. (2012)}]{ddm12} 
de Martino D. et al. 2012,  (in preparation)

\bibitem[{Falanga et al. (2010)}]{fal10} 
Falanga M. et al.  2010,  Proc. 8th INTEGRAL Workshop, Dublin

\bibitem[{Hill et al. (2011)}]{hill11} 
Hill A. et al. 2011, \mnras, 415, 235

\bibitem[{{Masetti} {et~al.}(2006)}]{mas06}
{Masetti} N. {et~al.}, 2006, \aap, 459, 21

\bibitem[{{Pretorius}(2009)}]{pre09}
{Pretorius} M. 2009, \mnras, 395, 386

\bibitem[{{Saitou} {et~al.}(2009)}]{sai09}
{Saitou} K. et al. 2009, \pasj, 61, 13

\bibitem[{Saitou et al. (2011)}]{sai11} 
Saitou K. et al. 2011, arxiv.org/abs/1105.4717

\bibitem[{Sazonov \& Revnivtsev (2004)}]{saz04} 
Sazonov S. \& Revnivtsev, A.\ M.,  2004, \aap, 423, 469

\bibitem[{Takata \& Taam (2009)}]{tak09} 
Takata J. \& Taam R., 2009, \apj, 702, 100

\bibitem[{Tam et al. (2010)}]{tam10} 
Tam P.  et al. 2010, \apj, 724, 207

\bibitem[{Thorstensen \& Amstrong (2005)}]{tho05} 
Thorstensen J. \& Amstrong E., 2005, \apj, 130, 759


\end{thebibliography}

\bigskip
\noindent {\bf DISCUSSION}

\bigskip
\noindent {\bf M. REVNIVTSEV:} First a comment : there is one more argument against CV identification.
 It is the absence of the thermal Fe line in X-rays. 
  Otherwise, do you see two-peak profiles in optical lines as expected from disk ?

\bigskip
\noindent {\bf J.M. BONNET-BIDAUD:} I agree with the absence of Fe line. 
No, we don't see two-peak profiles in optical lines but this may indicate 
an origin on the companion star and not in the disk.

\end{document}